# Investigation into the Itinerant metamagnetism of $Sr_3Ru_2O_7$ for the Field Parallel to the Ruthenium Oxygen Planes

Robin S. Perry[1,2*], Takashi Tayama[3], Kentaro Kitagawa[1], Toshiro Sakakibara[3], Kenji Ishida[2] and Yoshiteru Maeno[1,2]

[1] *Department of Physics, Kyoto University, Kyoto 606-8502, Japan*
[2] *Kyoto University International Innovation Center, Kyoto 606-8502, Japan*
[3] *Institute of Solid State Physics, The University of Tokyo, Kashiwa, Chiba 277-8581, Japan*

**Abstract**

We report a detailed investigation into the metamagnetism of $Sr_3Ru_2O_7$ at low temperatures for the magnetic field parallel to the ruthenium oxygen planes. The metamagnetism is studied as a function of temperature, magnetic field and sample quality using magnetisation, magnetotransport and specific heat as probes. From hysteretic behaviour in the magnetisation, we confirm earlier work and observe a finite temperature critical point at (5 T, >0.25 K). In our highest quality samples two-step metamagnetic transitions are additionally observed at 5.8 T and at 6.3 T, which coincide with a range of broad maximum in the magnetoresistance. At low temperatures, these two metamagnetic features each further split in two. Such behaviour of the multiple transitions are qualitatively different from the first order transition at 5.1 T.

\* Present address : School of Physics and Astronomy, University of St. Andrews, North Haugh, St. Andrews, Fife KY16 9SS, UK.

## 1. Introduction

Quantum criticality is currently generating large interest in the solid state physics community because of its potential to describe strongly correlated electron systems in a wide perspective.[1] A quantum critical point (QCP) exists when a continuous phase transition occurs at zero temperature and the properties of the material are then governed by quantum instead of thermal fluctuations. Instead of long wavelength correlations diverging near a classical critical point, both spatial and temporal correlations are expected to diverge near a QCP.[2] These fluctuations, associated with the Heisenberg uncertainty principle, can affect the properties over a large area of the phase diagram.[3] It is for these reasons that quantum criticality is being considered as a valid theoretical framework for such diverse areas of solid state physics as heavy fermions and cuprate high-transition-temperature superconductors. Recently, a new type of correlation that diverges only in the time domain has been proposed near a QCP.[4]

The study of metamagnetic quantum criticality is a relatively new field within the larger field of quantum criticality.[5] Metamagnetic quantum criticality occurs when a critical point terminating a line of first-order phase transitions in the field-temperature phase diagram is suppressed to zero temperature. This is of particular interest because the tuning parameter to access the quantum critical behaviour is magnetic field and not pressure. Thus many more experimental techniques are available for a detailed investigation of the system by fine-tuning the control parameter. The phenomenon of itinerant metamagnetism has been observed in many materials although $Sr_3Ru_2O_7$ was, to the best of our knowledge, the first material in which it was examined explicitly in terms of quantum criticality.[5] Since then metamagnetic QCPs have been noted in other systems.[6] Unusual behaviour in the scattering was observed in the proximity of the QCP in $Sr_3Ru_2O_7$ leading to the suggestion that a genuinely novel state is stabilised. This discovery was made possible due to an advance in crystal quality[7,8]; in older generation crystals the novel behaviour was masked by disorder. Here, we continue the investigation by making low-temperature magnetisation, resistivity and specific heat measurements for the field orientation perpendicular to the previous study. Our two major findings are, firstly, discovery of a family of metamagnetic transitions between 5.8 T and 6.3 T and, secondly, the confirmation of a finite-temperature critical-point at 5 T and ~0.2 K. In the narrow field range between 5.8 T and 6.3 T we observe enhanced scattering below ~0.8 K we use qualitative arguments to propose that this anomalous region about 6 T maybe caused by proximity to a metamagnetic QCP.

$Sr_3Ru_2O_7$ is the bilayered member of the perovskite Strontium Ruthenate series and is the sister compound to the well-known spin-triplet superconductor $Sr_2RuO_4$.[9] It is a Fermi liquid at low temperatures with a large electronic specific heat $\gamma$ of 110 mJ/Ru mol $K^2$ indicating strong electronic correlations. Its large Wilson ratio of 10 and structural similarity to the tri-layered, ferromagnetic $Sr_4Ru_3O_{10}$ have lead to the generally accepted view that this material is an exchange-enhanced paramagnet on the verge of ferromagnetism.[10] Generally, metamagnetism is empirically defined as a super-linear rise in the magnetisation $M$ as a function of magnetic field $B$ at a critical field $B_M$. The metamagnetism in $Sr_3Ru_2O_7$ has some anisotropy with $B_M$ =5 T for the field parallel to the ruthenium oxygen planes (denoted as the $B^{ab}$ orientation) and $B_M$ = 8 T for the field perpendicular to the *ab* plane (the $B^c$ orientation).[11] A second metamagnetic transition at 5.8 T for $B^{ab}$ was recently observed by Ohmichi and coworkers but they detected no first order phase transition down to 0.5 K for either field orientation.[12] Grigera and co-workers confirmed from the *ac* susceptibility measurements that $B_M$ evolves continuously as a function of angle between the two orientations.[13] More interestingly, they showed from features in the out-of-phase susceptibility, that a finite-temperature critical-point resides at 5 T and $T_{crit} \approx 1.2$ K for $B^{ab}$. This critical point decreases in temperature continuously as a function of field angle to the *c* axis to zero Kelvin at around 10°: the proposed metamagnetic QCP in the system. This unusual angle sensitive feature of the metamagnetism in $Sr_3Ru_2O_7$ gives us an excellent opportunity to study both the classical and quantum metamagnetic critical-points in the same system.

Crystal quality has been shown to be of great importance in this system. In crystals with an in-plane residual resistivity $\rho_{res}$=2.8 μΩcm a $T^3$ power law was observed at $B \approx B_M$ for $T$<0.5 K. Recently, we showed that in crystals with $\rho_{res}$=0.4 μΩcm the $T^3$ power law evolved into a minimum in $\rho(T)$ at ~1.1 K.[7] Here, we present the first low temperature investigation of the metamagnetism in this material for the $B^{ab}$ orientation in the new generation of high-quality single-crystals of $Sr_3Ru_2O_7$.[8] We found hysteretic behavior at the super-linear rise in the *M-H* curve, which is a clear indication of the first-order transition occurring. This is confirmation of Grigera's observation of a finite-temperature critical point by measurement using a static thermodynamic probe. We also found, in our highest quality samples, two-step metamagnetic transition at 5.8 T and 6.3 T, which coincide with a range of broad maximum in the magnetoresistance. Such behavior highlight the importance of clean samples for uncovering novel behaviour in correlated electron systems, especially in ruthenate compounds.

## 2. Experimental Method

All the crystals used in this study were grown by floating zone technique at Kyoto University and were characterized by Xray diffraction, resistivity and magnetisation (SQUID) measurements. Here, we define crystal quality in term of the residual or impurity resistivity $\rho_{res}$ for the current *I* parallel to the $RuO_2$ planes. $\rho_{res}$ was measured by fitting a $T^2$ Fermi liquid power law to the zero field resistivity between 7 K and 2 K. By optimizing the conditions of the crystal growth, we have recently succeeded in synthesizing ultra high quality crystals with $\rho_{res}$ =0.4 μΩcm which compares to the old generation of crystals with $\rho_{res}$ >2 μΩcm.[8] We refer to the new and old generation crystals as high quality and low quality, respectively. We present magnetisation, magnetotransport and specific heat data over a temperature range of 0.06 K to 10 K and for magnetic fields ($B^{ab}$ <12 T). Crystals of varying quality were measured and, where possible, we present the effects of impurities on the metamagnetism. The magnetisation was measured using a capacitance cell magnetometer situated at the Institute of Solid State Physics, the University of Tokyo. This apparatus allows us to measure the magnetisation in magnetic fields up to 14 T and at temperatures as low as 0.06 K.[14] The magnetoresistance was measured using a standard four probe technique with the current parallel to the *ab* plane and the magnetic field parallel to the current in the 'Lorentz force free' configuration. A dilution refrigerator situated at Kyoto University was used to access temperatures down to 0.06 K and magnetic fields up to 10 T. Lastly, the specific heat of a single crystal of mass 18 mg was measured using a Quantum Design PPMS system at Kyoto University. The error due to crystal misalignment was estimated to be less than 5° for each measurement.

## 3. Results and discussion

### 3.1 The effect of impurities on the metamagnetism

In Fig. 1(a), we plot the static magnetisation across $B_M$ for two crystals of different quality at low temperature. The effect of crystal quality on the magnetism in this material is clearly demonstrated: the magnetisation jump at 5 T is far sharper in the high quality crystal. There are also extra features resolved in the high quality crystals, which we shall return to later. At the lowest temperatures, small but clear hysteresis is resolved in the magnetisation (see inset to Fig. 1(a)). This represents direct evidence that there is indeed a first order phase transition for $B^{ab}$. We have tracked the hysteresis up to around 0.25 K where it falls below our resolution and the result was also confirmed in high quality crystals. Our results are consistent with the finding of Ohmichi *et al.* since the observed critical point is below their base temperature. However, there is a quantitative discrepancy in the magnitude of the temperature of critical point compared to the susceptibility study by Grigera and co-workers. We observe the hysteresis to vanish around 0.25 K suggesting that this is the critical temperature whereas Grigera *et al.* place the $T_{crit}$ at around 1.2 K from the dissipative component of the ac susceptiblity. Crystals of similar quality were used: Perry grew them both at Kyoto University.[8] The field resolution of our capacitance cell magnetometer used in the present study is limited to ~20 Oe by the magnet power supply. The large percentage errors on $\Delta B$ makes precise determination of $T_{crit}$ difficult although we can place a lower bound on $T_{crit}$ at 0.25 K. Such quantitative disagreement aside, our observation supports the claim that a finite temperature critical point resides in the $H-T$ phase diagram for this field orientation.

In Fig. 1(b), the low temperature magnetotransport is plotted as a function of magnetic field for two crystals of different quality. There is a dramatic difference between the magnetoresistance for the two crystals. In both crystals, there is a cusp in the magnetoresistance at the 5 T associated with the first order phase transition. The cusp in the higher quality crystal is better defined but the general form of the magnetoresistance remains unchanged across the 5 T transition for the two crystals. However, at around 6 T the high quality crystal displays a broad maximum not observed in the low quality crystal. This 'mesa' structure (see Fig. 3 for a large scale plot) is quite remarkable and highlights the importance of crystal quality in strongly correlated electron materials. The nature of the impurities in these crystals is unclear but due to the floating zone growth technique the impurities are likely to be ruthenium vacancies i.e. non-magnetic.[8] There has been little work done on the effect of impurities on metamagnetism in the literature. Although, one notable exception is the study with transition metals ions implanted into Palladium .[15] The effect of magnetic impurities was to push the system towards ferromagnetism indicated by lowering $B_M$. In $Sr_3Ru_2O_7$, the impurities appear have a very weak effect of $B_M$ supporting the picture that the residual scattering is dominated by non-magnetic ruthenium vacancies.

**3.2 High temperature investigation**

In order to investigate the critical behaviour further we have measured the magnetotransport, magnetisation and specific heat also at elevated temperatures. In particular we have focused on the unusual resistivity 'mesa' between 5.8 T and 6.4 T. In Fig. 2 we plot the static susceptibility $\chi(B)$ across the metamagnetic region as a function of magnetic field and temperature in our high quality crystals. $\chi(B)$ was obtained numerically by taking the field derivative of the magnetisation. The most striking feature of our data is the emergence of a rich structure of peaks in the 'mesa' field region at low temperatures. The second peak at 5.8 T has become a double peak structure and a third transition is resolved at 6.3 T. The split peaks of the 5.8 T and 6.3 T features are unlikely to be due to mis-aligned single crystal domains because we would expect the 5.0 T peak to be split as well. It is interesting that the edges of the magnetoresistance 'mesa' structure coincide with the two sets of the double peaks in $\chi(B)$. We observed no hysteresis in the magnetisation for these transitions although we cannot rule out the presence of first order phase transitions.

In Fig. 3, we plot the magnetoresistance $\rho(B)$ as a function of field and temperature for the high quality sample. The clear discontinuity in the gradient of $\rho(B)$ coinciding with the phase transition at 5 T is observed at all temperatures. In contrast, the mesa is highly temperature dependent. It evolves from a single peak above ~0.8 K to a broad maximum below ~0.8 K as the temperature is decreased. Also, at ~5.8 T the left-hand side of the mesa evolves into a double feature at ~0.6 K which is washed out by ~0.8 K. From Fig. 2, this is a similar temperature at which the double-peak in $\chi(5.8\ T)$ disappears. The double-peak in $\chi(6.3\ T)$ has a strong temperature dependence and is lost by 0.4 K coinciding with the disappearance of the right-hand side of the magnetoresistance mesa. These similar energy scales supports the suggestion that there is a connection between the mesa in the resistivity and the twin double-peaks in the susceptibility. Temperature sweeps of the resistivity also reveal interesting behaviour (see inset to Fig. 3). Below ~4.5 T and above ~7 T a Fermi liquid $T^2$ power law is observed as previously reported.[11] We observe an unusual kink in the resistivity at 6.0 T and 6.2 T and ~0.7 K (see arrow in inset to Fig. 3). This feature only exists in a field range within the mesa field range.

The specific heat divided by temperature versus temperature is plotted in Fig. 4. There are several interesting features to note. Firstly, there is broad maximum which shifts to lower temperature with increasing field: the peak temperature $T_{max}$ falls with increasing field until at 6 T it has gone completely. The second main point is that at 6 T $C/T$ increases nearly logarithmically down to 1 K, compared to $C/T$ at 5 T and 7 T which both saturate at low temperatures. A logarithmic temperature dependence of the electronic specific heat is expected in a non-Fermi liquid near a QCP.[16]

**3.3 Discussion**

Possibly the most interesting discovery of this study is the family of metamagnetic transitions around 6 T. Successive transitions at 60 T and 80 T have been observed in the metamagnet YCo$_3$.[17] In this material, the Co atoms are on different crystallographic sites and the two metamagnet transitions occur successively on different sites. Indeed, Ohmichi *et al.* discussed the second transition (in lower quality crystals) as possibility originating from the second crystallographic Ru site in the bilayered perovskite structure. However, since there are now three transitions, the underlying mechanism is unlikely to be due to a different local density of states at each Ru site. Ohmichi also proposed the idea of orbital dependent metamagnetism based the assumption that one can assign a dominant orbital character to each Fermi surface branch. Recent quantum oscillation measurements [7,18] and band structure calculations have uncovered multiple bands in Sr$_3$Ru$_2$O$_7$ but the Fermi surface appears to be complicated and it may be difficult to assign specific *d* orbitals to particular bands.[19] Multiple transitions may also be attributed to multiple peaks in the density of states with the same orbital character. We cannot confirm or deny either argument at this time.

It is interesting to examine the qualitative differences in the data between the 5 T transition and the 5.8 T and 6.3 T crossovers. A single peak in $\chi(B)$ and a cusp in $\rho(B)$ at 5 T can be compared to split peaks in $\chi(B)$ and the unusual 'mesa' in $\rho(B)$ between 5.8 T and 6.3 T. The 'mesa' structure could result from mixed phases being stabilized close to the first order phase transition and the presence of magnetic domains could explain the increase in the scattering around 6 T.[20] However, mixed phases are usually a simple consequence of demagnetisation altering the internal magnetic field of the sample and smearing the phase transition. This would cause a plateau in the susceptibility as a function of field, which is not observed. Also the demagnetising field is of the order 10$^{-2}$ T which is two orders of magnitude smaller than the applied field discounting the possibility of a mixed phase being present. Another possibility could be the presence of a metamagnetic quantum critical point (MMQCP) in the system. The MMQCP would be associated with the 5.8 T and 6.3 T transitions where the anomalous behaviour is observed. Support for this picture comes from the logarithmic divergence of the electronic specific heat at 6T down to 1 K. In this range of temperature we also see a resistance exponent of <2 at 6 T and following $\rho$ to lower temperature reveals the previously mentioned kink as the mesa region is entered.[21] We also cannot rule out the possibility that critical point sits at negative temperature. A final point is that the resistivity peak at ~6 T is reminiscent of the feature seen at 8 T for $B^c$. Both peaks have a similar general shape although the $B^{ab}$ peak is twice as wide in field and has a magnitude of 0.13 μΩcm compared to 1.21 μΩcm for $B^c$ and it is maybe possible that the underlying physics for the two phenomena is the same.

One of the best-studied itinerant metamagnets is CeRu$_2$Si$_2$.[22] In this material $B_M = 7.7$ T and no first order phase transition has been observed down to 0.05 K.[23] The temperature dependence of the specific heat has some similarities to that of Sr$_3$Ru$_2$O$_7$. Around $B_M$, there is a weak maximum in $C/T$ (percentage increase of $C/T$ between $T=0$ K and $T=T_{max}$ is $\Delta(C/T)_{max}$ ~4 % at 7 T) which decreases in temperature as $B \rightarrow B_M$. At all fields $C/T$ is temperature independent below some temperature reflecting that the system is Fermi liquid even at $B=B_M$. In comparison, Sr$_3$Ru$_2$O$_7$ is quite asymmetric between the high field and low field region and the percentage increase $\Delta(C/T)_{max}$ ~20% at 4 T. Aoki et al. showed that all of the features in the specific heat of $CeRu_2Si_2$ could be understood in terms of the Fermi level sitting close to a peak in the density of states. If this were the case in $Sr_3Ru_2O_7$ we would expect to observe the maximum in $C/T$ for $B>B_M$, contrary to the experimental result. The peak for $B<B_M$ may represent the energy scale for the crossover of the critical fluctuations from antiferromagnetic (low temperature) to ferromagnetic (high temperature). Neutron scattering measurements have observed ferromagnetic fluctuations at high temperatures (>25 K) and antiferromagnetic fluctuations in the ground state.[24] Nevertheless, the large Wilson ratio of 10 indicates that low **q** enhancements are important in this system. Borzi et al. reported that the changes of the quantum oscillation frequencies through $B_M$ in $Sr_3Ru_2O_7$ are not so large compared to $CeRu_2Si_2$,[18] in which the total Fermi surface volume is thought to change due to localization of the *f* electron [25, 26]. Since the change of Fermi surface topology is suggested to be quite small in $Sr_3Ru_2O_7$, the spin-fluctuation change at $B_M$ in $Sr_3Ru_2O_7$ might be different from that in $CeRu_2Si_2$. Therefore detailed neutron scattering measurements are highly desired to uncover the change of the spin-fluctuation character through metamagnetic transition, which are now in progress.

Finally, we evaluate the Kadowaki-Woods ratio (=$A/\gamma^2$ where $A$ is the coefficient of the $T^2$ scattering rate and $\gamma$ is the electronic density of states) and the Wilson ratio $R_W$ (=$7.3\times10^4\times\chi$ (emu/mol)/$\gamma$ (mJ/mol $K^2$)) at several fields across the metamagnetic transition (Table I). At all fields in Table I a $T^2$ scattering rate was observed and $A$ was extracted in a temperature range of 0.2 K to 0.6 K. The specific heat data at 1 K is used and although $\gamma$ is not temperature independent at this temperature for all fields in Table I the relative comparison of $R_W$ and the Kadowaki woods ratio is still useful. The universal value of Kadowaki-Woods ratio for heavy Fermion systems is $1\times10^{-5}$ $\mu\Omega cm(mole\ K/mJ)^2$ which is over an order of magnitude larger for *d* band metals.[27] The Kadowaki-Woods ratio doubles between the high and low field states. The *A* parameter is a function of the particle density $n$, the Fermi wave vector $k_F$ and the quasiparticle mass $m^*$. Thus, given that we know the Fermi surface changes across the transition, it is reasonable to assume the change in $A/\gamma^2$ may be due to the change of $k_F$. This would then be expected to drive a change in the quasiparticle mass and particle density. At low fields $R_W$=10 is in agreement with Ikeda and co-workers.[10] Across the transition $R_W$ drops to 8 at 7 T suggesting the ferromagnetic fluctuations are not further enhanced in the high-field region.

5. Summary and Conclusions

We have conducted an investigation into the metamagnetism in high quality crystals of $Sr_3Ru_2O_7$ for the magnetic field parallel to the $RuO_2$ planes. We present good evidence from hysteresis in the low temperature magnetisation for a finite-temperature critical point at 5 T and >0.2 K. This transition is also clearly observed in the resistivity. We have also uncovered multiple metamagnetic transitions between 5.8 T and 6.3 T in our highest quality crystals. These transitions appear to behave qualitatively differently from the 5 T transition and we have observed no first order phase transition. Anomalous behaviour between 5.8 T and 6.4 T includes enhanced scattering in a narrow field window bounded by split peaks in the static susceptibility. The resistivity as a function of temperature also displays an unusual kink at low temperatures. Although the data are inconclusive the rich structure we present here highlights the importance of clean samples for uncovering novel behaviour in correlated electron systems.

**Acknowledgements**

We acknowledge useful discussions with S. Nakatsuji, K. Ishida, H. Yaguchi, K. Kitagawa, S. A. Grigera, R. Borzi and A. P. Mackenzie. This work has been supported by Grants-in-Aid for Scientific Research from the Japan Society for Promotion of Science and from the Ministry of Education, Culture, Sports, Science, and Technology (MEXT). RSP is supported by a Grant-in-Aid for the 21st Century COE "Center for Diversity and Universality in Physics" from MEXT.

Table I. Fitting parameters extracted from resistivity, specific heat and magnetisation data for several fields across the metamagnetic transition of $Sr_3Ru_2O_7$. $A$ is the coefficient of the Fermi liquid $T^2$ scattering power, $\gamma$ is the specific heat at 1.0 K and $\chi$ is the static susceptibility at 0.06 K.

| $B^{ab}$ (T) | $A$ ($\mu\Omega$cm/K$^2$) | $\gamma$ (1 K) (mJ/Ru mol K$^2$) | $\chi$ (emu/Ru mol) | $A/\gamma^2$ ($\mu\Omega$cm (Ru mol K / mJ)$^2$) | $R_W$ |
|---|---|---|---|---|---|
| 1.0 | 0.077 | 107 | 0.015 | 6.7E-06 | 10 |
| 3.0 | 0.090 | 107 | 0.015 | 7.9E-06 | 10 |
| 4.0 | 0.111 | 116 | 0.018 | 9.1E-06 | 12 |
| 5.2 | 0.347 | 174 | 0.090 | 1.1E-05 | 37 |
| 7.0 | 0.259 | 139 | 0.016 | 1.3E-05 | 8 |

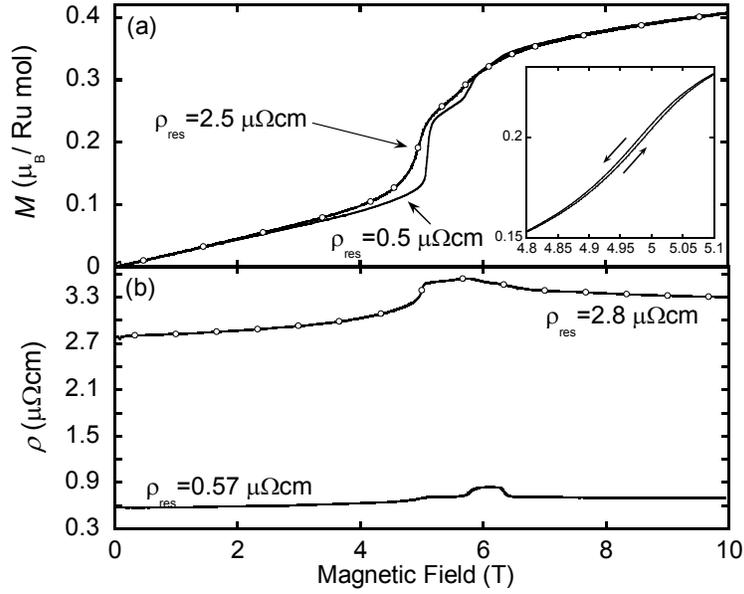

Figure 1. The effect of impurities on the metamagnetism in $Sr_3Ru_2O_7$ for $B^{ab}$. In (a) the magnetisation is plotted as a function of magnetic field for a high quality and a low quality crystal (open circles) at temperatures of 0.06 K and 0.07 K, respectively. The inset shows hysteresis in the magnetisation across the 5 T metamagnetic transition at 0.06 K for a low quality crystals. The arrows show the direction of the field sweep. In (b) the resistivity ratio versus magnetic field for high and low quality crystals at 0.1 K and 0.06 K, respectively.

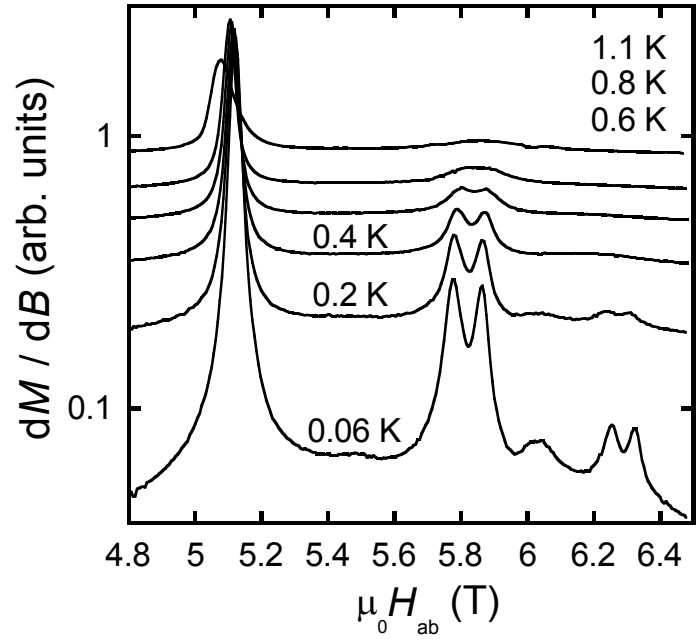

Figure 2. The static susceptibility as a function of magnetic field and temperature in high quality single crystals of $Sr_3Ru_2O_7$. For clarity, the data has a vertical offset that is linearly proportional to the temperature of measurement.

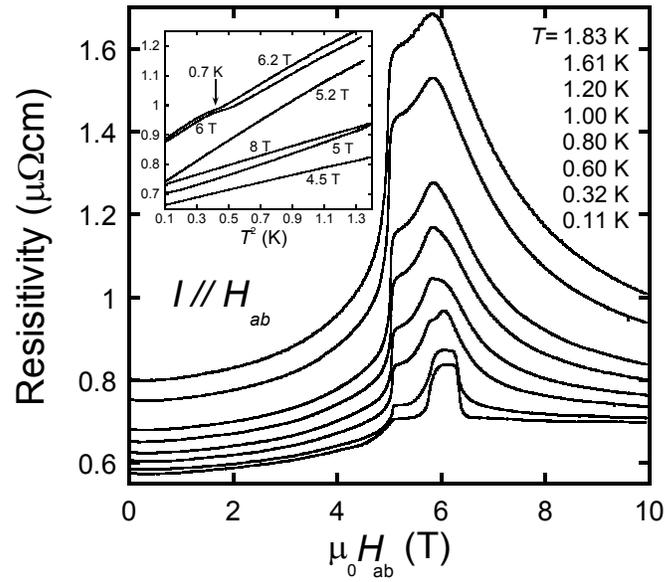

Figure 3. The magnetoresistance as a function of magnetic field and temperature for high quality crystals. The current $I$ is parallel to the $ab$ plane. The inset shows resistivity versus $T^2$ at fixed fields.

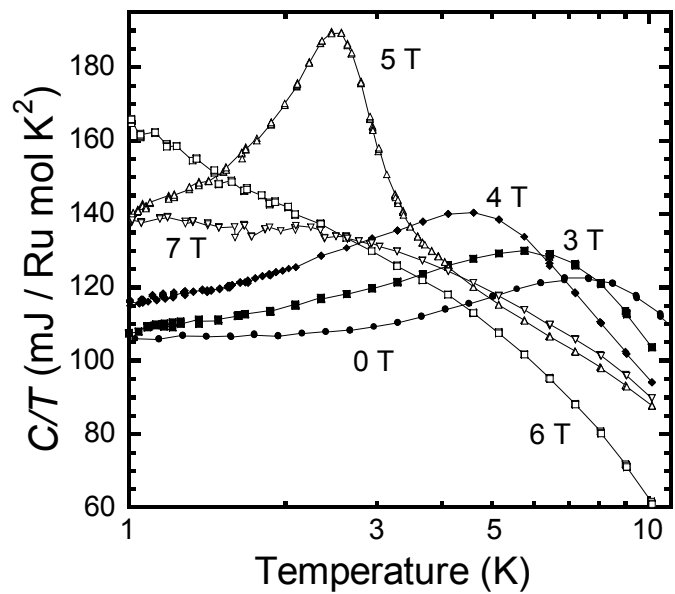

Figure 4. The specific heat divided by temperature of a high quality crystal of $Sr_3Ru_2O_7$ as a function of temperature and magnetic field. The phonon contribution (fitted between 20 K and 30 K) has been subtracted from the data.